\let\ni=\noindent

\documentstyle[aps,amssymb]{revtex}
\tightenlines
\begin{document}


\title{  Noether's theorem for the variational equations}

\author{ C. M. Arizmendi,$^{1}$ J. Delgado,$^2$ H. N. N\'u\~nez-Y\'epez,$^{3}$  A. L. Salas-Brito$^{4}$\footnote{Corresponding author}} 

\address{$^1$ Departamento de F\'{\i}sica, Facultad de Ingenier\'{\i}a,\\
Universidad Nacional de Mar del Plata, Mar del Plata, Argentina }

\address{$^{2}$ Departamento de Matem\'aticas, 
        Universidad Aut\'onoma Metropolitana-Iztapalapa, \\
         Apar\-tado Postal  55-534 Iztapalapa 09340 D.\ F., M\'exico.}

\address{$^{3}$ Departamento de F\'{\i}sica, 
        Universidad Aut\'onoma Metropolitana-Iztapalapa, \\
         Apar\-tado Postal  55-534 Iztapalapa 09340 D.\ F., M\'exico.}

\address{$^4$Laboratorio de Sistemas Din\'amicos,
Departamento de Ciencias B\'asicas,\\ Universidad
Aut\'onoma Metropolitana-Azcapotzalco,\\ Apartado
Postal 21-267,  Coyoac\'an 04000 D.\ F., M\'exico.}
\maketitle
\date{\today}

\begin{abstract}  We introduce an generalized action functional describing the equations of motion and the variational equations for any Lagrangian system. Using this novel scheme we are able to generalize Noether's theorem in such a way that to any $n$-parameter continuous symmetry group of the Lagrangian there exist 1) the usual $n$ constants of motion and 2) $n$ extra constants valid in the variational equations. The  new constants are related to the infinitesimal generators of the symmetry transformation by relations similar to the ones that stem from the `nonextended' Noether theorem.\end{abstract}

The equations of motion of many physically relevant systems can be obtained within a Lagrangian framework.  The systems are described using a  Lagrangian density through the Euler-Lagrange equations. To these equations of motion we can associate another set of equations obtained by linearizing the Euler-Lagrange  equations around one of its particular solutions. Such linearized equations are  called   the {\it variational equations} \cite{Whittaker}, are important in studies of stability \cite{arnold,lanczos,matsuno} and for defining  quantities as the Lyapunov exponents or some forms of entropy  in dynamical system theory\cite{arnold,jackson}. Also they are  the equations of geodesic deviation in general relativity and other metric theories of gravitation\cite{mtw,robin,pla94,sussman2,taub}. They are in general useful for describing  perturbations from the original dynamics and have been variously used in analysing non-linear  evolution equations \cite{case85}, in  studies of stability in galactic dynamics\cite{chandra}, in performing the Painlev\'e test\cite{steeb}, and  in  geometric control theory\cite{uam00,jurd}.  

It has been shown that if the field equations can be  casted in Hamiltonian form and if any solution of the variational equations can be expressed as the Poisson bracket of a field variable, $\phi_a$, with any other dynamical variable,  then this dynamical variable is necessarily  a constant of motion\cite{case85}. The converse of this result has been  proved earlier in \cite{case78}. The variational equations and its symmetry properties are also important for studying solitonic solutions to certain non-linear  wave equations\cite{matsuno}. It is also known that if the original Lagrangian is invariant under a translation in any space-time direction and $\psi$ is one of its solutions, then the directional derivative of $\psi$ along such  direction  is then a solution of the variational equations\cite{case78}.  

Thus, given the relevance of the variational equations in many aspects of physics, and of the relationship\cite{case85,case78} between constants of motion and properties of the solutions of the original equations, in this work we want to pursue   the relation between symmetries and constants of motion in the first-order dynamics.  To accomplish this task,  a Lagrangian description of the variational equations is very convenient. Though this description does not appear easy to attain---given the use of particular (and explicitly space-time dependent) solutions of the original equations for deriving the variational equations--- a complete Lagrangian characterization of the variational equations of Lagrangian systems has been advanced recently\cite{pla00}, see also \cite{robin,taub} for previous related results. The idea is the following, given the Lagrangian density associated with a system, $L$, a new  density  needs to be defined \cite{uam00,pla00} as

\begin{equation}\label{gamma}
\gamma= \frac {\partial L} {\partial \phi^a} \epsilon^a + \frac {\partial L} {\partial \phi^a_{,\mu}} \epsilon^a_{,\mu},
\end{equation}

\ni where the $\epsilon^a$ and the $\epsilon^a_{,\mu}$ correspond, respectively, to deviations from the original field variables and to their space-time derivatives. These deviations are assumed to connect two nearby solutions of the original Euler-Lagrange equations. In (\ref{gamma}), as in all of the paper, the summation convention is  implied, we also use  $\psi_{,\mu}\equiv\partial \psi/ \partial x^\mu$, Latin indices correspond to internal variables, and Greek indices to space-time coordinates: $x^\mu, \,\mu=0,\dots, 3$. From a strictly mathematical viewpoint the  density $\gamma$ can be  interpreted   as the prolongation of $L$\cite{uam00}. 

Using  the density $\gamma$ we can further define an alternative action functional as

\begin{equation}\label{action}
\Sigma= \int_{\Omega} \gamma(\phi^a, \epsilon^a, \phi^a_{,\mu}, \epsilon^a_{,\mu})\, d^4x, 
\end{equation}

\ni where $\Omega$ is an appropriate space-time region. This action is fundamental for our formulation.

 The extremalization of (\ref{action}) gives  \cite{pla00} both the Euler-Lagrange equations,

\begin{equation}\label{ELE}
\left(\frac {\partial L} {\partial \phi^a_{,\mu}}\right)_{,\mu} - \frac {\partial L} {\partial \phi^a} =0,
\end{equation}

\ni   and the associated variational  equations,

\begin{equation}\label{VE}
\left(\frac{\partial^2 L}{ \partial  \phi^a_{,\mu} \partial 
\phi^b_{,\nu}}\right) \epsilon^b_{,\mu \nu} + \left[ 
   \left( \frac{\partial^2L}{ \partial \phi^a_{,\mu} \partial \phi^b_{,\nu}}
      \right)_{,\mu} + \frac{\partial^2 L}{ \partial  \phi^a_{,\nu} \partial \phi^b} -
	\frac{\partial^2 L}{ \partial \phi^b_{,\nu} \partial \phi^a}\right]\epsilon^b_{,\nu}  + \left[ \left( \frac{\partial^2L}{ \partial\phi^a_{,\mu} \phi^b} \right)_{,\mu}  -
\frac{\partial^2 L}{ \partial 
 \phi^a \partial \phi^b}\right]\epsilon^b=0. 
\end{equation}

  In the form given above, the variational equations do not have to be regarded as  explicitly dependent on the space-time coordinates unless the original Lagrangian density $L$  is so from the start. This  property, shared by the function $\gamma$,  is analogous to what occurs with the Lagrangian density $L$ itself,  which is not considered  space-time dependent despite being  an explicit function of the fields $\phi^a(x)$ which, in principle,  also depend on $x^\mu$.   The form (\ref{VE}) of the variational equations and their origin from  (\ref{action}) is closely connected with properties of the Jacobi equation of interest for the study of geodesics in Riemannian manifolds and gives a way for deriving the associated curvature tensor from a variational formulation \cite{arnold,amici}. It should be clear that $\gamma$ plays the role of a new form of the  density associated with the system.  From the existence of the action (\ref{action}) it can be easily proved that $\gamma$ is invariant under arbitrary point transformations  on its configuration manifold comprised by the $\phi^a$, the $\epsilon^a$, and their space-time derivatives\cite{uam00,pla00}. The function  $\gamma$ thus describes both the system and its variational equations.

To establish the connection between symmetries and conservation laws in the variational equations, let us consider a prolonged Lagrangian density $\gamma$ and assume it  invariant (or quasi-invariant, but in what follows we use invariant for short) under the following continuous group of transformations 

\begin{eqnarray}\label{groupeq}
 \bar{\phi}^a&=& F^a_{\bar{\phi}}(x^\mu,\phi^a,\epsilon^a,w^s),\\ 
\bar{\epsilon}^\mu&=&F^\mu_{\bar{\epsilon}}(x^\mu,\phi^a,\epsilon^a,w^s),
\\
\bar{x}^\mu&=&F^\mu_{\bar{x}}(x^\mu,\phi^a,\epsilon^a,w^s),
 \end{eqnarray}

\ni where the $w^s$ are the  $r$ parameters of the group, defined in such a way that the transformation reduces to the identity when all the $w^s$ vanish.  The infinitesimal generators of the transformation group (\ref{groupeq})  are thus

\begin{equation}\label{igtg}
\zeta^a_{s}= \left(\frac{\partial F^a_{\bar{\phi}}}{\partial w^s}\right)_{w^s=0},\qquad
\eta^a_{s}= \left(\frac{\partial F^a_{\bar{\epsilon}}}{\partial w^s}\right)_{w^s=0},\qquad
\xi^\mu_s= \left(\frac {\partial F^\mu_{\bar{x}}} {\partial w^s}\right)_{w^s=0}.
\end{equation}

\ni   If the set of transformations (\ref{groupeq}--7) [or (\ref{igtg})] is a symmetry  of $\gamma$, it could stem directly  from a symmetry of the original Lagrangian $L$ and, in this case, we could  have $F^a_{\bar{\phi}}=F^\mu_{\bar{\epsilon}}$. But it is also possible  that this will not be the case, for the  symmetries of $\gamma$ are expected to be larger than those of $L$ ---as can be simply ascertained  by thinking about the possible  transformations  that leave (\ref{ELE}) and (\ref{VE}) unchanged. Our  examples below show simple instances of such case. 

 That the transformation  group (\ref{groupeq}) be a symmetry of $\gamma$,  means that the action $\Sigma$ (Eq.\ \ref{action})  remains invariant under the transformations generated by the group. A necessary and sufficient condition for this to be true is that

\begin{equation} \label{consym}
\gamma \left(\bar{\phi}^a, \frac {\partial\bar{\phi}^a}{\partial\bar{x}^\mu},\bar{x}^\mu\right) D  =\gamma(\phi^a,\frac{\partial\phi^a}{\partial x^\mu},x^\mu).
\end{equation}

\ni where $D=\det\left( {\partial \bar{x}^\mu}/ {\partial x^\nu}\right)$ is the Jacobian determinant of the transformation. Eq.\ (\ref{consym}) is the fundamental equation allowing us to explicitly write the relationship between the generators of the continuous symmetries of $\gamma$  and the conserved quantities in the variational equations\cite{rund}. 

To see how the extension can be accomplished, differentiate partially this equality (\ref{consym}) with respect to $w^s$,  noting that the right hand side is  independent of such parameters, and evaluate in  $w^s=0$ after the differentiation. After some manipulations this yields

\begin{eqnarray}
\left[\gamma\xi^\mu_s + \frac{\partial \gamma}{\partial \phi^a_{,\mu}}\left(\zeta^a_s - \phi^a_{,\nu}\xi^\nu_s\right)+ \frac{\partial \gamma}{\partial \epsilon^a_{,\mu}}\left(\eta^a_s - \epsilon^a_{,\nu}\xi^\nu_s\right)\right]_{,\mu}&&+
\left[ \frac{\partial \gamma}{\partial \phi^a}-\left(\frac{\partial \gamma}{\partial \phi^a_{,\mu}}\right)_{,\mu}   \right]\left(\zeta^a_s - \phi^a_{,\nu}\xi^\nu_s\right)\nonumber\\
&&+\left[ \frac{\partial \gamma}{\partial \epsilon^a}-\left(\frac{\partial \gamma}{\partial \epsilon^a_{,\mu}}\right)_{,\mu}   \right]\left(\eta^a_s - \epsilon^a_{,\nu}\xi^\nu_s\right)=0.
\end{eqnarray}

\ni  Finally,   use the equations of motion [Eqs. (\ref{ELE}) and (\ref{VE})] to obtain 

\begin{equation} \label{nt}
 \frac {\partial \tau^\mu_s} {\partial x^\mu}=0
\end{equation}

\noindent where the divergenceless tensor $ \tau^\mu_s$ is defined by

\begin{equation}\label{emt}
\tau^\mu_s= \gamma\xi^\mu_s + \frac{\partial \gamma}{\partial \phi^a_{,\mu}}\left(\zeta^a_s - \phi^a_{,\nu}\xi^\nu_s\right) + \frac{\partial \gamma}{\partial \epsilon^a_{,\mu}}\left(\eta^a_s - \epsilon^a_{,\nu}\xi^\nu_s\right).
\end{equation}

\noindent This  is the conserved tensor associated with the variational equations.  It is to be noted that this result can be regarded as the Noether's theorem for the variational equations  ---that can be checked following the steps for the demonstration of Noether's theorem that can be found in many places, for example in \cite{rund}. The conservation of $\tau$ (\ref{emt}) was to be expected  since $\gamma$ is indeed a Lagrangian function on its own.   The conserved quantities are simply related to the infinitesimal generators (\ref{igtg}) of the symmetry group  of the variational equations, as  (\ref{emt}) explicitly shows. Notice that our results also show that in any Lagrangian system invariant under a $n$-parameter continuous group of transformations there exist $2n$ conserved quantities, the usual $n$ Noether conserved quantities plus the $n$ additional ones associated with the variational equations. Of course that not very one of the $n$ new quantities has to be  independent of the $n$ previously  known ones.

To give  examples of the usefulness of this result, we next derive  (in admittedly simplistic instances) conserved quantities  starting from  well-known Lagrangian symmetries. 

\noindent {\sl Example 1.} Let us consider the Lagrangian density

\begin{equation}\label{einstein}
L=\frac {1}{2}R^{a \mu b \nu}x_{a,\mu}x_{b,\nu},
\end{equation}

\ni  where $R^{a \mu b \nu}=R^{ b \nu a \mu}$ is the Riemann tensor. 

This Lagrangian   (\ref{einstein}) describes Einstein equations in a vacuum. The density $\gamma$ associated with (\ref{einstein}) is  

 \begin{equation}
\gamma= \frac {1} {2} \left(  R^{a \mu b \nu}  \right)_{,\rho}\epsilon^{\rho} x_{a,\mu} x_{b,\nu}+ R^{a \mu b \nu} x_{a,\mu} \epsilon_{b,\nu};
\end{equation}
 
\ni in this case $\gamma$ describes linearized gravitation in a vacuum. Taking into consideration the properties of $R^{a \mu b \nu}$,   it is not difficult to realize that  $\gamma$ is  invariant under the one parameter group of transformations 

\begin{equation}
\bar{x}^\mu =  x^\mu +x^\mu w, \quad \hbox{and} \quad \bar{\epsilon}^\mu=\epsilon^\mu,
\end{equation}

\ni the infinitesimal  generators are thus $\xi=0$, $\zeta^\mu=x^\mu$,  and $\eta^\mu=0$. 
With the symmetries and the infinitesimal generators established, the extended Noether's theorem predicts that the quantity

\begin{eqnarray}
&=&\frac{\partial\gamma}{\partial x^a_{,\mu}} \zeta^a\nonumber\\
        &=&\left(R^{a\mu b\nu}\right)_{,c}\epsilon_c\, x_{b,\nu}\, x_a +  R^{a\mu b\nu}\epsilon_{b,\nu}\, x_a\\
&=& R^{a\mu b\nu}\epsilon_{b,\nu}\, x_a
\end{eqnarray}

\ni  is divergenceless and hence generates a conserved quantity. This is  a rather useful tensor that was employed (and had to be evaluated using direct  calculations) in \cite{glass70} to calculate the conserved quantities of Newman and Penrose. Notice also that this $\tau^\mu$ is a conserved quantity  in {\sl any} vacuum spacetime in General Relativity. We have thus managed to derive using our result (\ref{emt}) one of the results in \cite{robin}. Moreover, $\gamma$ can be also shown to be invariant under the one parameter group of transformations

\begin{equation}\label{es}
\bar{x}^\mu =  x^\mu , \quad \hbox{and} \quad \bar{\epsilon}^\mu=\epsilon^\mu+\epsilon^\mu w,
\end{equation}

\ni with the  infinitesimal generators $\xi=0$, $\zeta^\mu=0$,  and $\eta^\mu=\epsilon^\mu$. In this case (\ref{emt}) leads directly to the divergence less tensor

\begin{eqnarray}\label{theta}
\theta^\mu&=&\frac{\partial\gamma}{\partial \epsilon^a_{,\mu}} \eta^a\nonumber\\
        &=& R^{a\mu b\nu}\epsilon_a\, x_{b,\nu},
\end{eqnarray}

\ni  which due to the symmetry properties of $R^{a\mu b\nu}$, happens to identically vanish. Nevertheless, we have illustrated in a simple instance that the symmetries of $\gamma$ can be larger than those of $L$.
  It should be noticed that the `extended' symmetry [equations (\ref{es})] is the original symmetry of the Lagrangian just applied  to the perturbation variables.  We must point out also that the idea for this example is  taken  from  \cite{robin} where it is used to relate symmetry transformations to the existence of  divergence less quantities in a first-order approximation to Einstein theory.

 There are other potential uses of our main result (\ref{nt}) in general relativity and in other field theories. For example, for investigating perturbations to known metrics \cite{ernst} and to contribute to the study of certain angular momentum ambiguities of recent interest \cite{tichy}.  The excellent review  \cite{ellis} discusses some other contemporary uses of perturbation methods in cosmological models in the context of general relativity.  

\ni {\sl Example 2.} Our result can be also profitably applied to particle mechanics by a simple reinterpretation of the symbols used in Eq.\ (\ref{emt}) and the replacement of the four spacetime parameters $x^a$ by the time $t$. To give a simple example, let us consider   a Lagrangian function describing a system of particles in which one of the coordinates is ignorable. In this case, the original Lagrangian and the function $\gamma$ can be written as

\begin{eqnarray}\label{class}
L&=&\frac {1} {2} m_{ab} \dot q^a \dot q^b -U(q),\\ 
\gamma &=&  m_{ab}\dot q^a\dot \epsilon^b -\frac {\partial U}{\partial q^a}\epsilon^a ,
\end{eqnarray}

\ni where $m_{ab}$ is a symmetric `mass' matrix formed by the second derivatives  of the system's  kinetic energy respect to the generalized velocities $\dot q^a$, and $U(q)$ is its potential energy function. For the purposes of the example we are assuming that $ m_{ab}$ does not depend on the generalized coordinates $q^a$, but this is of no important for the argument ---it just simplifies a little the form of the associated conserved quantity. The invariance under translations in the, let us say, $q^A$ direction ({\it i. e.} the transformation is $\bar{q}^A=q^A+w $, $\bar{\epsilon}^A=\epsilon^A+w $, the rest of the coordinates and velocities remain unchanged) can be associated with the infinitesimal generators $\xi=0$ and $\zeta^A=\eta^A=1$, and so using  (\ref{emt}) we can  obtain that 

 \begin{eqnarray}\label{newp}
\pi_A&=& \frac {\partial \gamma} {\partial \dot \epsilon^A}+ \frac {\partial \gamma} {\partial \dot q^A}\nonumber\\
     &=&  p_A+\frac {\partial \gamma} {\partial \dot q^A},
\end{eqnarray}

\ni is a constant of motion in the variational equations of the system, where $p_A=\partial L/\partial \dot q_A$ is the  momentum conjugate to the coordinate $q^A$. In fact, using the obvious invariance of (\ref{class}) under the space translation $\bar{q}^A=q^A+w $, and $\bar{\epsilon}^A=\epsilon^A+w $, which corresponds to the infinitesimal generators $\zeta^A=1$, $\eta^A=1$, and $\xi=0$, we  can directly prove that  the term 

\begin{equation}
\frac {\partial \gamma} {\partial \dot q^A}
\end{equation}

\ni appearing in Eq.\ (\ref{newp}), is, as it should be,  a constant of motion on its own. These  results  are  direct and simple examples of  conserved quantities which may be important (in more complex instances) in the analysis of classical perturbation methods with  astronomical interest \cite{orbits}.

To summarize, we have established a form of Noether's theorem that encompasses the variational equations and  have illustrated its use with  direct examples. The conserved quantities encountered can be  useful in perturbation theory as our formalism can be regarded as a starting point for studying   perturbations  in all kind of Lagrangian problems. Moreover, if we take into consideration  the Lagrangian foundations of the path integral approach  our formalism could be useful in approaching approximate schemes in quantum field theory\cite{weinberg}. We also consider it as possibly important for  studying properties  of solitonic solutions in nonlinear equations \cite{matsuno}.  Besides, the formalism could have  some bearings to  modern mathematical developments in  Lagrangian field theories\cite{sarda}. 

\acknowledgements { }

This work was partially supported by  the  Universidad Nacional Aut\'onoma de M\'exico under  a PAPIIT-IN grant. We  thank  D. C. Robinson (King's College, London), R. Sussman (ICN-UNAM) and R. P. Mart\'{\i}nez-y-Romero (FC-UNAM) for their helpful remarks or suggestions.  We also thank  H. Schwartzi, M. S. Salas-N\'u\~nez, and  P. M. Zura and all her gang for lots of cheerful enthusiasm.  A.L.S.-B. acknowledges  the hospitality   of the Department of Physics of Emory University where parts of this work were carried out.  
This work is dedicated to the loving memory of F. C. Bonito (1987--2002).

\end{document}